# OPTICS AND PHOTONICS

Mr. Thomas Büttner
Prof. Ben Eggleton
School of Physics, A28
The University of Sydney, NSW 2006
Australia
thomasb@physics.usyd.edu.au  Phone: (T.B.): +61 293 513 963
egg@physics.usyd.edu.au      Phone: (B.E.): +61 293 513 604

## Phase-locking in Multi-Frequency Brillouin Oscillator via Four Wave Mixing

Thomas F. S. Büttner[1], Irina V. Kabakova[1], Darren D. Hudson[1], Ravi Pant[1], Christopher G. Poulton[1,2], Alexander C. Judge[1], and Benjamin J. Eggleton[1]

[1]Centre for Ultrahigh bandwidth Devices for Optical Systems (CUDOS), Institute of Photonics and Optical Science (IPOS), School of Physics, University of Sydney, NSW, 2006, Australia

[2]CUDOS, School of Mathematical Sciences, University of Technology, Sydney, NSW, 2007, Australia

**Stimulated Brillouin scattering (SBS) and Kerr-nonlinear four wave-mixing (FWM) are among the most important and widely studied nonlinear effects in optical fibres. At high powers SBS can be cascaded producing multiple Stokes waves spaced by the Brillouin frequency shift. Here, we investigate the complex nonlinear interaction of the cascade of Stokes waves, generated in a Fabry-Perot chalcogenide fibre resonator through the combined action of SBS and FWM. We demonstrate the existence of parameter regimes, in which pump and Stokes waves attain a phase-locked steady state. Real-time measurements of 40ps pulses with 8GHz repetition rate are presented, confirming short-and long-term stability. Numerical simulations qualitatively agree with experiments and show the significance of FWM in phase-locking of pump and Stokes waves. Our findings can be applied for the design of novel picosecond pulse sources with GHz repetition rate for optical communication systems.**

Stimulated Brillouin Scattering (SBS), whereby light interacts coherently with acoustic phonons, is one of the strongest nonlinear effects observed in optical fibres[1-3]. In the SBS process, a narrow-band pump field of frequency $\omega_0$ can generate a strong backward-propagating Stokes field of frequency $\omega_1 = \omega_0 - \Omega_B$ through interaction with acoustic phonons (frequency $\Omega_B$) once a certain power threshold is reached. Although SBS in optical fibres can be detrimental to optical systems, it has also enabled a range of important technologies such as Brillouin fibre lasers[4,5], sensors[6], pulse compressors[7], phase conjugators[8], as well as devices for slow light[9], stored light[10] and all-optical signal processing[11,12].

Brillouin lasing is achieved by generating Stokes photons via SBS in a cavity[4]. Due to the cavity feedback, the power threshold for the generation of the Stokes waves is reduced considerably[13]. The linewidth of the Stokes field can then be much narrower than the Brillouin gain bandwidth $\Delta \nu_B = 1/(2\pi\tau)$, where $\tau$ is the phonon lifetime[3]. In Brillouin lasers with short cavities, i.e. cavities short enough such that the free spectral range of the cavity is of the order of $\Delta \nu_B$, only one cavity mode experiences strong Brillouin gain. Such lasers emit a single frequency Stokes wave once the lasing threshold is reached. In the general (off-resonant) case, the peak of the SBS gain spectrum $(\omega_0 - \Omega_B)$ does not coincide with a cavity resonance and the frequency of the Stokes wave $\omega_1$ is determined by an interplay between Brillouin gain, nonlinear phase-shift from the Kerr-nonlinearity, and cavity selectivity, known as Stokes detuning[14].

SBS can be cascaded in resonators, which produces multiple Stokes waves spaced by $\Omega_B$ thus forming a Brillouin frequency comb[15]. The acoustic waves that couple adjacent comb components have slightly different propagation constants[14,16]. Due to phase-matching conditions, coupling of more than two optical waves with a common acoustic wave is only possible over distances of a few millimeters[17]. SBS provides no inherent mechanism to produce or stabilise any particular phase-relationship between the comb components. In the absence of any additional nonlinear effect, each Stokes wave grows from noise thus possesses a random phase upon generation[18]. In addition, phase-noise and unequally spaced comb components, due to different detunings of the Stokes waves, lead to a changing spectral phase with time. The lack of coherence between spectral lines prevents the generation of pulse trains via the SBS process alone.

Different cavity configurations have been suggested to achieve phase-locking of Brillouin frequency combs[17,19–21]. Experimentally, phase-locking of Brillouin frequency combs has been demonstrated by achieving coupling of all Stokes waves with a common acoustic wave by limiting the SBS interaction to a short part of the cavity[20] and by using modal dispersion in a short-multimode fibre[21]. Recently, an autocorrelation measurement of a Brillouin comb generated in a Brillouin-erbium fibre laser has been presented that suggests pulse-like behaviour in the time domain[22]. However, the phase-locking mechanism for this configuration has not been explained and the autocorrelation presented is not unambiguous evidence of a pulse train due to a relatively large autocorrelation background[23].

Kerr-induced four wave-mixing (FWM) also leads to generation of new frequencies and, unlike SBS, is known to couple three or more optical waves in a phase-sensitive manner[3,24]. It has been shown that frequency combs generated through FWM in high-finesse micro-resonators can exhibit certain spectral phase signatures[25].

In resonators, SBS and FWM can efficiently co-exist. Degenerate FWM between co-propagating pump and Stokes waves (generated via SBS) create Anti-Stokes waves at frequencies $\omega_j = \omega_0 - j \times \Omega_B$, for $j < 0$, and higher-order Stokes waves at frequencies $\omega_j = \omega_0 - j \times \Omega_B$ for $j \geq 2$. Generation of Anti-Stokes waves is a clear sign of the presence of FWM since these waves cannot be generated by SBS. The Stokes waves generated via FWM, in turn, can act as a seed and can be further amplified by SBS, thus reducing the threshold for higher-order Stokes waves[26–29].

In this work, we demonstrate the generation of a phase-locked Brillouin frequency comb with repeatable spectral phase via the interplay of SBS and FWM. Real-time measurements of the comb show stable picosecond pulse trains with GHz repetition rate. Our experiment was performed in a short ($\sim 38\,\text{cm}$), low finesse, Fabry-Perot fibre cavity. A numerical study of the system shows good agreement with experimental results and clearly establishes that FWM is essential for the system's dynamics to attain a phase-locked steady state.

This new understanding of the interaction between FWM and SBS can potentially be exploited for creating novel picosecond pulse sources with GHz repetition rate for optical communication systems and high speed optical clocks[30].

**Results**

Brillouin frequency combs were generated by coupling quasi-continuous waves (quasi-CW) pump light into a Fabry-Perot cavity consisting of a short piece of $As_2Se_3$ chalcogenide glass[31] fibre as illustrated schematically in Fig. 1. Quasi-CW light was used to avoid thermal effects by keeping the average power low ($\sim 10\,\text{mW}$) while obtaining high peak powers ($\sim 1\,\text{W}$). The pump light consisted of narrow band $500\,\text{ns}$ square pulses at $1550\,\text{nm}$. The polarisation of the pump was adjusted such that only one polarisation mode was excited. Due to a large refractive index of $2.81$, the Fresnel reflections on the perpendicularly cleaved fibre facets provided feedback of about $R = 22.6\%$. A single transverse mode with an effective area of $A_{\text{eff}} = 56\,\mu\text{m}^2$ was guided by the fiber. The large Brillouin gain[32,33] $g_B = 6.10 \times 10^{-9}\,\text{m/W}$ and nonlinear refractive index[34] $n_2 = 2.4 \times 10^{-13}\,\text{cm}^2/\text{W}$ of the chalcogenide glass allowed us to perform the experiment in a short fiber length $L = 38.31\,\text{cm}$. The phonon lifetime in $As_2Se_3$ is $\tau = 12\,\text{ns}$ and the Brillouin frequency shift was measured to be $\Omega_B/(2\pi) = 7.805\,\text{GHz}$. The linear propagation loss of the $As_2Se_3$ fibre[32] is $\alpha = 0.84\,\text{dB/m}$ and the dispersion coefficient[31] is about $-504\,\text{ps}\,\text{km}^{-1}\text{nm}^{-1}$. Details about the experimental setup can be found in the Methods section.

Important parameters of the system dynamics[35] are the linear phase-shift $\Delta\varphi_j$, of the pump $(j=0)$ and Stokes waves $(j=1,2,..)$ per cavity roundtrip. It is convenient be describe these linear phase shifts in the form

$$\Delta\varphi_j = \Delta\varphi_0 - j\beta\,, \qquad (1)$$

where $\Delta\varphi_0 = 2\omega_0 nL/c$ is the linear phase shift of the pump per round trip and $\beta = 2\Omega_B nL/c$ is the difference of the linear phase shifts of the two adjacent comb components per round trip (without Stokes detuning). Here, we neglected the dispersion of the effective index $n$ and the Brillouin shift $\Omega_B$ because we only consider a small frequency range (~30 GHz). Since $\Delta\varphi_0$ represents an absolute phase, it is sensitive to small changes of $n$ and $L$ due to temperature drift of the cavity and also to frequency tuning of the pump laser. In the experiment $\Delta\varphi_0$ has not been stabilised and its values are unknown. The parameter $\beta$ is a phase-difference and is far less sensitive to pump tuning and temperature drift than $\Delta\varphi_0$ and is assumed to be constant in the parameter range considered here. The parameter $\beta$ can be controlled by cleaving the fibre to a certain length since $\partial\beta/\partial L = 2\Omega_B n/c = 2\pi/(6.84\,\text{mm})$. For both, experiments and simulations, $\beta$ was chosen to be a multiple of $2\pi$. However, we found that this is not the only condition for observing phase-locking and qualitatively similar results were also obtained in experiments and simulations for different values of $\beta$.

**Experiment**

Figure 2(a) shows the optical spectrum of the input light. The spectrum consists of a single line at $\lambda = 1550.2\,\text{nm}$ corresponding to the pump light (P). A time domain measurement of the input light is plotted in the inset in Fig. 2(a), showing a square pulse of $500\,\text{ns}$ length with a slight slope decreasing toward the end of the pulse arising from the amplification process. A typical spectrum measured at the output of the chalcogenide fiber for about 0.7 W peak power coupled into the fibre is plotted in Fig. 2(b). Besides the pump light (P) the output spectrum also exhibits four Stokes lines at longer wavelength (1S-4S) and two anti-Stokes waves (1AS, 2AS) at shorter wavelengths compared to the pump (P). The Stokes waves (1S-4S) were generated through the interplay of SBS and FWM and the wavelength (1AS, 2AS) were generated through FWM of pump and Stokes waves[13,36]. The spacing between two neighboring lines is about $\Delta\lambda = 63\,\text{pm}$, which is equal to the SBS frequency shift of the As$_2$Se fibre $\Delta\lambda = \Omega_B \lambda^2 / (2\pi c)$ at $\lambda = 1550\,\text{nm}$, where $c$ is the vacuum speed of light.

Two real-time measurements of output light from the chalcogenide fibre cavity are shown in Fig. 3(a) and 3(b) for the input powers of about 0.7 W but for different values of $\Delta\varphi_0$. The insets to (a) and (b) show zoomed-in sections of 1 ns time intervals of the main graph. Compared to the input quasi-CW pulses, the output pulses show rapid temporal modulations after about 25 ns which are not resolved in the main graph. The inset at about $30\,\text{ns}$ in Fig. 3(a) reveals a cosine modulation, which we attribute to beating between the pump and the newly generated first Stokes wave. The frequency of the beat signal corresponds to $\Omega_B / (2\pi)$, *i.e.* the Brillouin frequency shift.

We first consider envelopes of the interference signals in Fig. 3(a) and 3(b). In Fig. 3(a) the envelope of the fast modulation is oscillating with a period of about $100\,\text{ns}$, whereas no modulation can be seen in Fig. 3(b). The two insets (at $246\,\text{ns}$ and $373\,\text{ns}$) in Fig. 3(a) reveal a strong second frequency of $2 \cdot \Omega_B / (2\pi)$. The spectrum shown in Fig. 2(b) suggests that the main contribution to this higher frequency is a result of interference between the pump and second-order Stokes wave. It is clear from the insets that the interference signal of the multiple waves in Fig. 3(a) is not constant throughout the $500\,\text{ns}$ pump duration and that the phase-relationship of the waves drifts in time (at the rate of about $1/100\,\text{ns} = 10\,\text{MHz}$). This drift arises from non-equally-spaced comb lines resulting from different Stokes detunings of the first and second-order Stokes waves.

Figure 3(b) shows a qualitatively different result obtained for a different value of $\Delta\varphi_0$. After about 100 ns the system reaches a steady state with a constant envelope of the modulation. The zoomed-in sections of the interference signal, shown in the insets at $200\,\text{ns}$ and $400\,\text{ns}$, reveal a stable train of ~40 ps sub-pulses with a repetition rate of ~ $\Omega_B/(2\pi)$. The pulse shape is the result of the interference of at least 3 waves (such as pump, first and second-order Stokes wave) equally spaced in frequency and with a constant phase relationship.

This phase-locked train of pulses is repeatable and stable if the experiment is repeated with the same experimental conditions. We demonstrated this by recording the output pulses for a series of input pump pulses.

Figure 4(a) shows 11 output pulses observed for 11 independent $500\,\text{ns}$ input pulses coupled into the fiber with $0.5\,\text{ms}$ delay between the pulses. All output pulses display very similar temporal behavior.

Figure 4(b) shows a 0.5 ns long zoomed-in section of the output pulses at $325\,\text{ns}$, revealing the same interference of the waves for all traces. From this we conclude that the waves that contribute the interference have the same phase-relationship for all measurements.

**Numerical Study**

The experiment indicates that a stable train of pulses can be obtained with a phase-relationship of the contributing waves, which is repeatable and constant in time. As outlined above, SBS alone does not possess a mechanism to provide either of these properties.

We believe that FWM can explain phase-locking with a repeatable phase-relationship. We demonstrated this by performing a numerical study of the system with and without FWM and compared the simulation results. For the sake of simplicity we only present results for the smallest number of waves that are necessary to qualitatively reproduce the experimental results in this paper and to demonstrate the phase-locking mechanism.

The model includes forward and backward propagating pump, first and second-order Stokes waves, and the corresponding four acoustic waves which couple these waves in a fibre oriented along the $z$-direction. The interaction between these six optical and four acoustic waves can be described in terms of the dynamic coupled mode-equations[14] (S3) – (S7) shown in the Supplementary Information. In these equations we included terms relevant to the SBS interaction, as well as the effects arising from the Kerr-nonlinearity: self-phase modulation (SPM), cross-phase modulation (XPM) and FWM. Simulation parameters were chosen to match experimental conditions.

To compare simulation and experimental results we look at the optical power at the output of the fibre. The optical (pump ($j=0$), first ($j=1$) and second ($j=2$) Stokes) waves in the fibre resonator can be described in terms of the complex amplitudes $E_j^\pm(z,t)$ of their electric fields:

$$E_j^\pm(z,t) = A_j^\pm(z,t)e^{-i\omega_j(t \mp nz/c)} = \sqrt{P_j^\pm(z,t)}e^{i\theta_j^\pm(z,t)}; \quad j=0,1,2. \qquad (2)$$

Here "$\pm$" indicates the direction of propagation, $A_j^\pm(z,t)$ are the slowly varying complex envelopes of the amplitudes $E_j^\pm(z,t)$. $\theta_j^\pm(z,t) = \varphi_j^\pm(z,t) - \omega_j(t \mp nz/c)$ represents the phases of the waves, $n \approx 2.81$ is the effective index, and $\omega_j$ are the steady state frequencies of the waves. We assume that the complex amplitudes $E_j^\pm(z,t)$ are normalized such that $P_j^\pm(z,t)$ represent the optical powers of the waves measured in dimensions of Watts. The output power $P_{out}(t)$ at the end of the fibre resonator can be calculated with

$$P_{out}(t) = (1-R)\,|\,E_0^+(L,t) + E_1^+(L,t) + E_2^+(L,t)\,|^2 \ . \qquad (3)$$

In order to obtain stable pulses at the output of the fibre, the powers $P_j^\pm(z,t)$ and the phases $\varphi_j^\pm(z,t)$ must attain a steady state. The total output power $P_{out}(t)$ in the steady state can be calculated by inserting equation (2) into equation (3) to yield

$$\begin{aligned}P_{out}(t)/(1-R) =& P_0^+(L)+P_1^+(L)+P_2^+(L)+2\sqrt{P_0^+(L)P_1^+(L)}\cos\left[(\omega_0-\omega_1)(t-t_0)\right]\\ &+2\sqrt{P_1^+(L)P_2^+(L)}\cos\left[\vartheta^+(L,t)+(\omega_0-\omega_1)(t-t_0)\right]\\ &+2\sqrt{P_0^+(L)P_2^+(L)}\cos\left[\vartheta^+(L,t)+2(\omega_0-\omega_1)(t-t_0)\right],\end{aligned} \quad (4)$$

where $t_0 = \left[\varphi_0^+(L)-\varphi_1^+(L)\right]/(\omega_0-\omega_1) \pm nz/c$ is a constant time offset. In Equation (4) we have introduced the phase parameter

$$\begin{aligned}\vartheta^\pm(z,t) &= \theta_0^\pm(z,t)-2\theta_1^\pm(z,t)+\theta_2^\pm(z,t)\\ &= \varphi_0^\pm(z,t)-2\varphi_1^\pm(z,t)+\varphi_2^\pm(z,t)-\left(t\mp\frac{nz}{c}\right)\left[(\omega_0-\omega_1)-(\omega_1-\omega_2)\right].\end{aligned} \quad (5)$$

$\vartheta^\pm(z,t)$ encapsulate the phase relationships between the three forward and backward propagating electric fields $E_1^+(z,t), E_2^+(z,t), E_3^+(z,t)$ and $E_1^-(z,t), E_2^-(z,t), E_3^-(z,t)$, respectively, which are critical for the phase-sensitive FWM interactions[24]. Equation (5) shows that even when the powers $P_j^+(L,t)$ and the phases $\varphi_j^+(L,t)$ are independent of time, $\vartheta^+(L,t)$ drifts with a constant rate if the frequencies $\omega_j$ are not equally spaced, preventing a stable interference signal $P_{out}(t)$. The steady state frequencies $\omega_1$ and $\omega_2$ of the Stokes waves are determined by a complicated interplay of Brillouin gain, nonlinear phase-shift from the Kerr-nonlinearity and cavity selectivity. In general, different order Stokes waves experience different amounts of Stokes detuning and the frequencies of the three waves are not equally spaced. Additionally, in order to obtain short pulses (of length $\sim 2\pi/(N\Omega_B)$, where $N$ is the number of participating waves[19]), $\vartheta^+(L,t)$ must also be close to zero to achieve the appropriate phasing of the three waves.

To illustrate the effect of FWM we first performed the simulations with and without FWM interaction while the terms for SBS, SPM and XPM remained in the equations.

Figure 5(a) and 5(b) show two qualitatively different results that were obtained by including FWM in the simulation, consistent with experimental conditions. Both simulations were performed using the estimated coupled input power of the experiment $P_{in} = 0.7\,\text{W}$. For the parameter $\Delta\varphi_0$ we choose $\Delta\varphi_0 = 0.548\,\pi$ and $\Delta\varphi_0 = 1.2\,\pi$, respectively, in order to obtain results that are qualitatively similar to the experimental results shown in Fig. 3(a) and 3(b). The plots on the left show the evolution of the powers $P_j^+(L,t)$ and the phase parameter $\vartheta^+(L,t)$ whereas the plots on the right display the output power of the fibre $P_{out}(t)$. The insets of the plots on the right show 1 ns long zoomed-in sections at different times. In Fig. 5(a) ($\Delta\varphi_0 = 1.2\,\pi$) neither the powers $P_j^+(L,t)$ nor the parameter $\vartheta^+(L,t)$ reach a steady state so the envelope of the burst is modulated and the pulses are not stable. However, for $\Delta\varphi_0 = 0.548\,\pi$ in Fig. 5(b) the phase parameter $\vartheta^+(L,t)$ and the powers $P_j^+(L,t)$ reach a steady state after an initial power transfer from pump to the Stokes waves. Since the powers $P_j^+(L,t)$ and $\vartheta^+(L,t)$ are independent of time, a stable train of pulses can be obtained.

Figure 5(c) shows simulation results obtained when FWM was not included in the simulation. The same parameters were used that generated the results of Fig. 5(b) ($P_{in}$ =0.7W, $\Delta\varphi_0 = 0.548\,\pi$). After an initial power transfer to the Stokes waves, the powers $P_j^+(L,t)$ of all waves reach a steady state. However, $\vartheta^+(L,t)$ drifts at a constant rate, which is the result of unequally spaced frequencies. Since $\vartheta^+(L,t)$ is not constant the interference signal changes with time and has a modulated envelope (Fig. 5(c), right).

The results of the simulation demonstrate that FWM links the parameter $\vartheta^{\pm}(z,t)$ to the powers $P_j^{\pm}(z,t)$ and *vice versa*. This important contribution of FWM is also illustrated by the steady state equations (S10) - (S15) shown in the Supplementary Information. The FWM terms (and only these terms) contain the phase parameters $\vartheta^\pm(z,t)$. From the steady state equations it follows that $\vartheta^\pm(z,t)$ must be time-independent, leading to the condition of phase-locking between the three waves. Furthermore, FWM couples the steady state powers to the phases and also the phases to each other via terms containing the phase parameters $\vartheta^\pm(z,t)$. In the absence of FWM the steady state powers $P_j^\pm(z,t)$ are independent of the phases as in a simple Brillouin laser configuration[14]. Thus, FWM is crucial to establishing a phase-locked output in the steady state.

In order to generalize these simulation results for a larger parameter space, we performed a two dimensional parameter scan. We chose to scan the input power $P_{in}$ and the parameter $\Delta\varphi_0$ since these parameters are controllable in the experiment for a given fibre sample by tuning the pump frequency and power. The input power $P_{in}$ was scanned in the interval from 0.4 W to 0.8 W in steps of 0.025 W and the phase shift $\Delta\varphi_0$ from 0 to $2\pi$ in steps of $0.05\pi$. For each set of parameters the coupled mode equations were integrated and an analysis (see Methods section for more details) was performed to determine whether the powers $P_j^\pm(z,t)$ and the parameter $\vartheta^+(L,t)$ reach a steady state.

The result of this parameter scan is shown in Fig. 6. The black area corresponds to the case where the phase parameter $\vartheta^+(L,t)$ and the powers $P_j^+(L)$ reach a steady state corresponding to phase-locking and resulting in a stable interference signal at the output of the fibre. These results qualitatively correspond to the result shown in Fig. 5(b). For the white area in Fig. 6, the parameter $\vartheta^+(L,t)$ does not reach a stable state and no stable interference signals $P_{out}^+(L,t)$ is obtained (similar to the result shown in Fig. 5(a)).

The same parameter scan was also performed for using the coupled mode equations without the FWM terms. We observed that without FWM there is no regime where the parameter $\vartheta^+(L,t)$ reaches a steady state.

**Discussion**

In this paper, we demonstrated phase-locking of Brillouin frequency combs generated by CW pumping of a short, low finesse Fabry-Perot resonator (Fig. 2(b)). By repeating the experiment with the same initial conditions we have shown that a mechanism exists that phase-locks the Brillouin frequency comb with a particular phase-relationship leading to the generation of stable trains of pulses (Fig. 4). Numerical simulations that included FWM showed a strong qualitative agreement with the experimental results and revealed that FWM is essential for phase-locking and the generation of stable pulse trains.

In developing the model we assumed negligible dispersion. The assumption was made since the linear contribution to phase-mismatch between pump and Stokes wave (arising from the dispersion of the effective index $n$) is much smaller than the nonlinear contributions due to nonlinear phase-shifts from SPM, XPM, FWM and SBS. This assumption is justified since the cavity length $L$ is much shorter than the coherence length of the degenerate FWM process[24] (~4 km) that is calculated using the material dispersion[31] of $As_2Se_3$.

In the current configuration the frequency detunings of the pump and Stokes waves with respect to the cavity resonances (described by the parameter $\Delta\varphi_0$) are sensitive to temperature drift of the resonator leading to a varying output of the resonator on a time scale of a few seconds. Besides active stabilization[37], reducing the cavity length and therefore increasing the FSR of the cavity could be used to improve stability. Recently, SBS has been demonstrated on photonic chips made of chalcogenide glass exploiting large SBS gain as well as strong mode confinement[38-40]. CW generation of several orders of Stokes waves has been demonstrated on-chip in a low finesse Fabry-Perot cavity of only a few cm length[13]. On-chip generation of phase-locked pulse trains via SBS and FWM promises a much higher degree of stability as well as more compact devices and will be the subject of future investigations.

**Methods**

A schematic of the experimental setup used is shown in Fig. 2(a). In order to perform the experiment in a quasi CW regime, the output of a continuous wave laser operating at 1550 nm with a linewidth of 500 kHz was carved into 500ns square pulses at 20 kHz repetition rate using an optical intensity modulator with extinction ratio of ~27 dB. This increased the linewidth of the pump to about 1.8 MHz. The modulator was followed by an erbium doped fibre amplifier (EDFA) to increase the peak intensity of the pulses. Modulating the CW light allowed to perform the experiment with high peak powers (~1 W) while keeping the coupled average power low (~10 mW). A polarisation controller was placed after the modulator to align the polarisation of the pump light with a polarisation mode of the chalcogenide fibre resonator. This was achieved by setting the pump power just above the threshold for the generation of the first-order Stokes wave and then adjusting the polarisation such that the power of the first-order Stokes wave is maximised. We used a 38.31 cm long $As_2Se_3$ step-index fibre (CorActive Inc.) with core and cladding diameters of ~5.25 μm and ~170 μm, respectively. The numerical aperture of the fibre was NA=0.17. Light was coupled into and out of the $As_2Se_3$ fibre by butt-coupling high numerical aperture silica fibres. An optical spectrum analyser (resolution ~10 pm) allowed spectral characterization of the output and a photodiode (bandwidth ~50 GHz) connected to a real-time oscilloscope (80 GSa/s, bandwidth 33 GHz) enabled real-time analysis of the transmitted signal. The real-time oscilloscope also allowed the recording of 110 consecutive pulse traces. In Fig. 4(a) and 4(b) every tenth trace of such a measurement is plotted.

Details about the coupled mode equations used for the numerical study can be found in the Supplementary Information. For the numerical integration of the time dependent couple mode equations we transformed the variables into frames moving with forward and backward propagating optical fields, respectively,[16,41] and used the integration method described by de Sterke et al.[41]. Steady states of the phase parameter $\vartheta^+(L,t)$ and the powers $P_j^+(L,t)$ were determined by integrating the coupled mode equations for a time interval of $6\,\mu s$ and calculating the standard deviations $\sigma(\vartheta^+)$ and $\sigma(P_j^+)$ of $\vartheta^+(L,t)$ and $P_j^+(L,t)$, respectively, in the time interval $[5\,\mu s, 6\,\mu s]$. $\vartheta^+(L,t)$ was considered to have reached a steady state if $\sigma(\vartheta^+) < 10^{-3} \times 2\pi$. The power $P_j^+(L,t)$ were considered to have reached a steady state if $\sigma(P_j^+) < 10^{-2} \times \langle P_j^+ \rangle$, where $\langle P_j^+ \rangle$ are the mean values of $P_j^+(L,t)$ in the interval $[5\,\mu s, 6\,\mu s]$.

**Acknowledgments**

The authors thank Prof. Martijn de Sterke for fruitful discussions. Funding from the Australian Research Council (ARC) through its Laureate Project FL120100029 is gratefully acknowledged. This research was also supported by the ARC Center of Excellence for Ultrahigh bandwidth Devices for Optical Systems (project number CE110001018). D. Hudson acknowledges support from an ARC Discovery Early Career Research Award (DE130101033).

**Author Contributions**

B.J.E., T.F.S.B., I.V.K. and R.P. conceived the experiment. T.F.S.B. and I.V.K. carried out the experiment. T.F.S.B., I.V.K., A.C.J. and C.G.P. discussed the theory and the simulation method. T.F.S.B. carried out the numerical simulations. T.F.S.B. wrote the manuscript and prepared the figures. B.J.E., I.V.K., A.C.J., C.G.P. and D.D.H. edited the manuscript. B.J.E., D.D.H. and I.V.K. supervised the project. All authors discussed the results and commented on the manuscript.

**Additional Information**

**Competing Financial Interests**

The authors declare no competing financial interests.

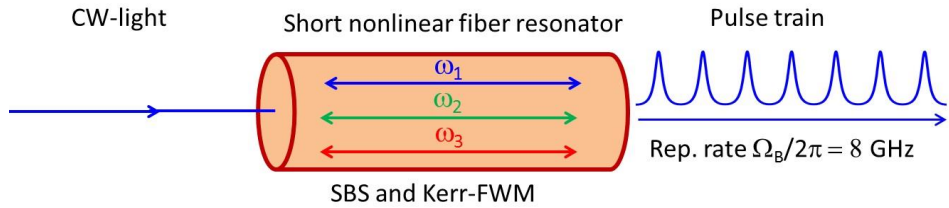

**Figure 1**| Concept of pulse train generation via SBS and FWM. CW light is coupled into a short, nonlinear fibre resonator. New frequencies are generated via SBS and FWM. At the output a train of phase-locked pulses can be observed with repetition rate equal to the SBS frequency shift.

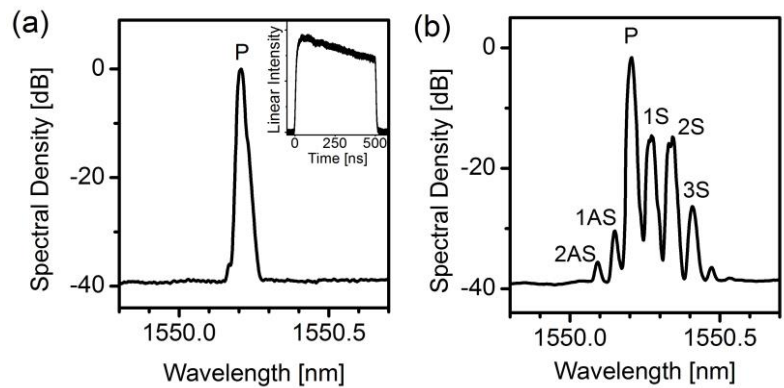

**Figure 2** | Optical spectra of input and output light of the fibre resonator. (a) Spectrum of the input light consisting of a single pump frequency (P). The inset shows the temporal measurement of a 500 ns input pulse. (b) Averaged output spectrum of the fibre resonator for about 0.7 W input peak power. Besides the pump (P), Stokes waves (1S-3S) at longer wavelength as well as anti-Stokes waves (1AS, 2AS) at shorter wavelengths are visible.

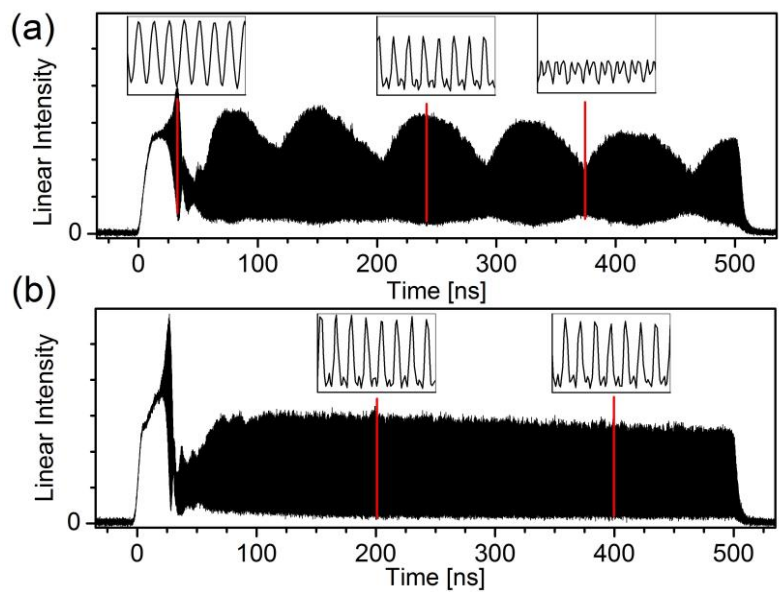

**Figure 3** | Temporal measurement of two output pulses of the fibre resonator for 500 ns input pulses. The insets show zoomed in sections of 1 ns length at different times. Both measurements were taken for similar input powers of ~0.7 W.

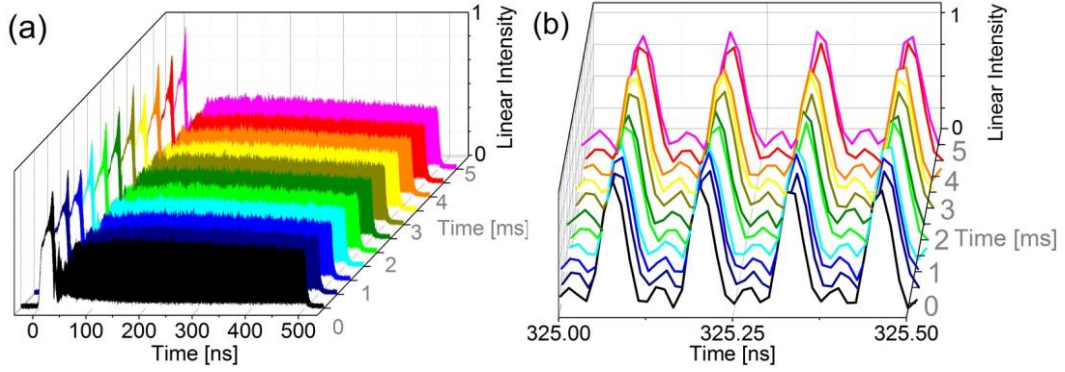

**Figure 4** | Temporal measurements of a series of eleven independent output pulses recorded in a 5 ms time interval with 0.5 ms time separation between the measurements. (a) Showing the entire 500 ns output pulses. (b) 0.5 ns long zoomed-in section of the pulses shown in (a) at 325 ns. (The maxima are aligned for better visibility.)

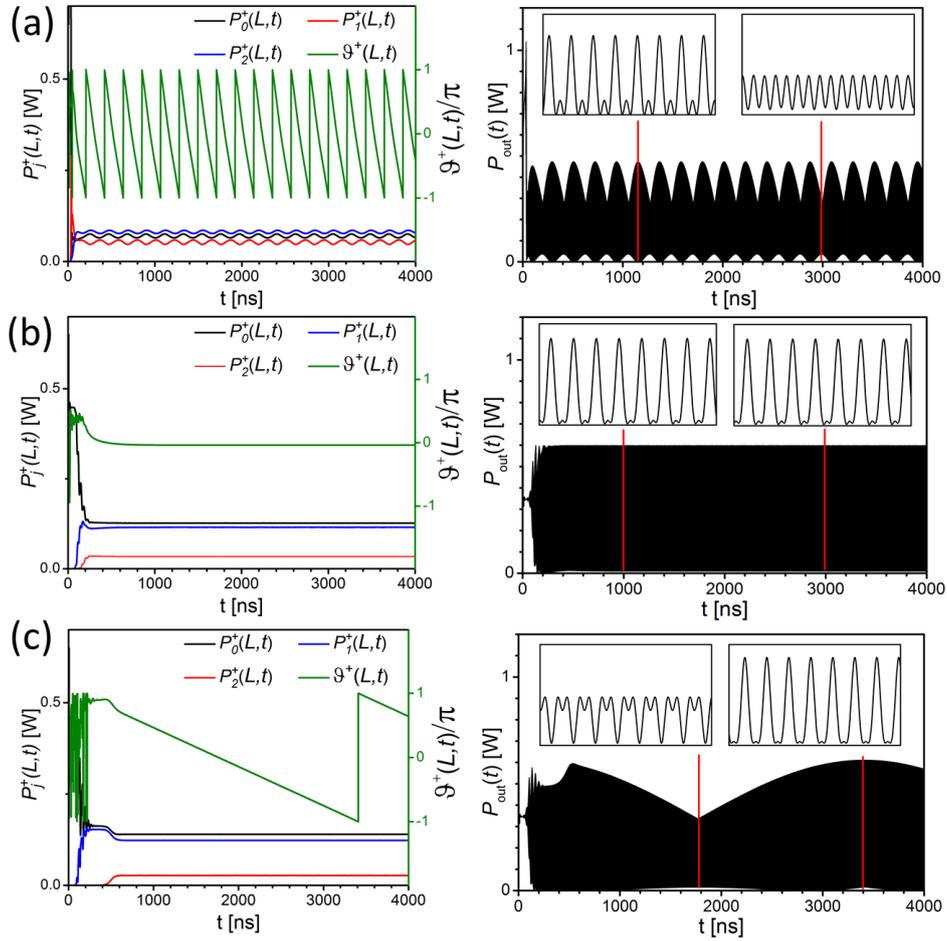

**Figure 5** | Left: Computed temporal evolution of the powers of pump, first and second-order Stokes wave $P_0^+(L,t), P_1^+(L,t), P_2^+(L,t)$, and the phase parameter $\vartheta^+(L,t)$ at the end of the fibre for input power $P_{in} = 0.7\,\text{W}$. Right: Total output power $P_{out}(t)$. (a) FWM included in the simulation, $\Delta\varphi_0 = 1.2\,\pi$. (b) FWM included in the simulation, $\Delta\varphi_0 = 0.548\,\pi$. (c) FWM not included in the simulation, $\Delta\varphi_0 = 0.548\,\pi$.

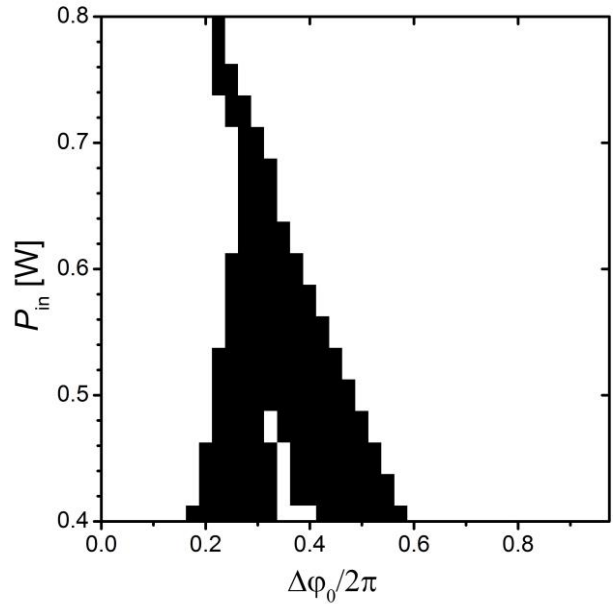

**Figure 6 |** Domains of phase-locking ($\partial \vartheta^+(L,t)/\partial t = 0$) (black) and drifting phase parameter ($\partial \vartheta^+(L,t)/\partial t \neq 0$) (white) for different input powers $P_{in}$ and phase shifts $\Delta\varphi_0$.

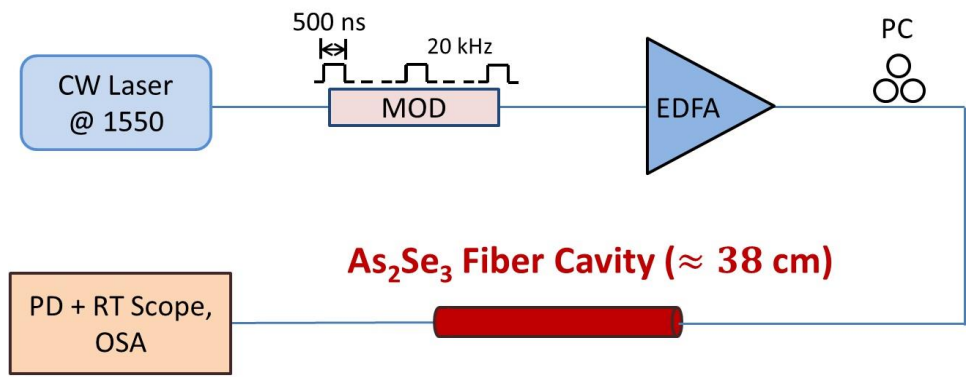

**Figure 7 |** Schematic of the experimental Setup. MOD: Optical intensity modulator. EDFA: Erbium-doped fibre amplifier. PC: Polarization controller. PD: Photodiode. RT-Scope: Real-time oscilloscope. OSA: Optical spectrum analyser.

# Supplementary Information for: Phase-locking in Multi-Frequency Brillouin Oscillator via Four Wave Mixing

Thomas F. S. Büttner[1], Irina V. Kabakova[1], Darren D. Hudson[1], Ravi Pant[1], Christopher G. Poulton[1,2], Alexander C. Judge[1], and Benjamin J. Eggleton[1]

[1]Centre for Ultrahigh bandwidth Devices for Optical Systems (CUDOS), Institute of Photonics and Optical Science (IPOS), School of Physics, University of Sydney, NSW, 2006, Australia

[2]CUDOS, School of Mathematical Sciences, University of Technology, Sydney, NSW, 2007, Australia

05.02.2014

Here, the numerical model used to obtain the simulation results presented in the main document is explained in detail. We also present the steady state equations for the optical powers of pump, first and second-order Stokes wave $P_0^\pm(z), P_1^\pm(z), P_2^\pm(z)$ and the phases $\varphi_1^\pm(z), \varphi_2^\pm(z), \varphi_3^\pm(z)$ of their slowly varying envelopes, allowing stable detuning of the two Stokes waves.

**Numerical Model**

Consistent with the experiment, we consider a short, low finesse, single-transverse mode, Fabry-Perot fibre resonator of length $L = 38.31\,\text{cm}$. The resonator is oriented along the $z$-axis and has a homogeneous filling with SBS gain constant[1] $g_B = 6.1 \times 10^{-9}\,\text{m/W}$ and nonlinear refractive index[2] $n_2 = 2.4 \times 10^{-13}\,\text{cm}^2/\text{W}$. Both facets of the resonator consist of reflecting surfaces with identical reflectivity $R \approx 22.6\%$. The linear loss of the fiber is $\alpha = 0.84\,\text{dB/m}$. We include pump (wavelength $\lambda = 1550\,\text{nm}$), 1st and 2nd Stokes waves in the model, which is the simplest configuration that displays the phase-locking mechanism of interest. Forward and backward propagating pump, first and second Stokes waves are represented by the complex amplitude of their electric fields $E_0^\pm(z,t)$, $E_1^\pm(z,t)$ and $E_2^\pm(z,t)$, respectively. The superscripts "$\pm$" represent the direction of propagation. The six electric fields are coupled by four acoustic fields represented by the complex amplitudes $M_1^\pm(z,t)$ and $M_2^\pm(z,t)$. The electric and acoustic fields can be described as

$$E_j^\pm(z,t) = \tilde{A}_j^\pm(z,t) e^{-i\tilde{\omega}_j(t \mp nz/c)}; \quad j = 0,1,2,$$
$$M_j^\pm(z,t) = \tilde{Q}_j^\pm(z,t) e^{-i[(\tilde{\omega}_{j-1} - \tilde{\omega}_j)t \mp (\tilde{\omega}_{j-1} + \tilde{\omega}_j)z]}; \quad j = 1,2, \quad \text{(S1)}$$

where $n = 2.81$ is the effective refractive index, and $\tilde{A}_j^\pm(z,t) = |\tilde{A}_j^\pm(z,t)| \exp(\tilde{\varphi}_j^\pm(z,t))$ are the slowly varying complex envelopes. For the frequencies $\tilde{\omega}_1$ and $\tilde{\omega}_2$ we use the ansatz: $\tilde{\omega}_j = \omega_0 - j\Omega_B$ (j=1,2), where $\omega_0 = 2\pi c/\lambda$ is the angular frequency of the pump, $c$ is the vacuum speed of light and $\Omega_B/2\pi = 7.805\,\text{GHz}$ is the Brillouin frequency shift.

Note, that we use a different definition for the frequencies $\tilde{\omega}_j$, the complex amplitudes $\tilde{A}_j^\pm(z,t)$ and the phases $\tilde{\varphi}_j^\pm(z,t)$ compared to the frequencies $\omega_j$, the complex amplitudes $A_j^\pm(z,t)$ and the phases $\varphi_j^\pm(z,t)$ referred to in the main document. The steady state frequencies of the Stokes waves $\omega_1$ and $\omega_2$ are unknown before the simulation reaches a steady state[3]. In the simulation, Stokes detuning manifests in global, constant phase drifts $\Delta\omega_j = -\partial\tilde{\varphi}_j^\pm(z,t)/\partial t$. When the simulation reaches a steady state, $\omega_j, A_j^\pm(z,t), \varphi_j^\pm(z,t)$ can be calculated with

$$\omega_j = \tilde{\omega}_j + \Delta\omega_j,$$
$$A_j^\pm(z,t) = \tilde{A}_j^\pm(z,t) e^{i\Delta\omega_j(t \mp nz/c)}, \quad \text{(S2)}$$
$$\varphi_j^\pm(z,t) = \tilde{\varphi}_j^\pm(z,t) + \Delta\omega_j(t \mp nz/c),$$

for $j = 0,1,2$. We set $\Delta\omega_0 = 0$, since the pump does not experience detuning.

The coupled mode equations[3,4] used to simulate the dynamic interaction between the six optical and four acoustic waves are

$$\pm\frac{\partial \tilde{A}_0^\pm}{\partial z} + \frac{n}{c}\frac{\partial \tilde{A}_0^\pm}{\partial t} = -\frac{\alpha}{2}\tilde{A}_0^\pm - \frac{g_B}{2A_{\text{eff}}}\tilde{A}_1^\mp \tilde{Q}_1^\pm + i\gamma\left(2P_T - P_0^\pm\right)\tilde{A}_0^\pm + i\gamma\tilde{A}_1^\pm\tilde{A}_1^\pm\tilde{A}_2^{\pm*}, \quad \text{(S3)}$$

$$\pm\frac{\partial \tilde{A}_1^{\pm}}{\partial z}+\frac{n}{c}\frac{\partial \tilde{A}_1^{\pm}}{\partial t}=-\frac{\alpha}{2}\tilde{A}_1^{\pm}-\frac{g_B}{2A_{\text{eff}}}\tilde{A}_2^{\mp}\tilde{Q}_2^{\pm}+\frac{g_B}{2A_{\text{eff}}}\tilde{A}_0^{\mp}\tilde{Q}_1^{\mp*}+i\gamma\left(2P_T-P_1^{\pm}\right)\tilde{A}_1^{\pm}+2i\gamma\tilde{A}_0^{\pm}\tilde{A}_2^{\pm}\tilde{A}_1^{\pm*},$$

(S4)

$$\pm\frac{\partial \tilde{A}_2^{\pm}}{\partial z}+\frac{n}{c}\frac{\partial \tilde{A}_2^{\pm}}{\partial t}=-\frac{\alpha}{2}\tilde{A}_2^{\pm}+\frac{g_B}{2A_{\text{eff}}}\tilde{A}_1^{\mp}\tilde{Q}_2^{\mp*}+i\gamma\left(2P_T-P_2^{\pm}\right)\tilde{A}_2^{\pm}+i\gamma\tilde{A}_1^{\pm}\tilde{A}_1^{\pm}\tilde{A}_0^{\pm*},\tag{S5}$$

$$\frac{\partial \tilde{Q}_1^{\pm}}{\partial t}=\frac{1}{2\tau}\left(\tilde{A}_0^{\pm}\tilde{A}_1^{\mp*}-\tilde{Q}_1^{\pm}\right)+f_1^{\pm},\tag{S6}$$

$$\frac{\partial \tilde{Q}_2^{\pm}}{\partial t}=\frac{1}{2\tau}\left(\tilde{A}_1^{\pm}\tilde{A}_2^{\mp*}-\tilde{Q}_2^{\pm}\right)+f_2^{\pm},\tag{S7}$$

where $A_{\text{eff}}=56\,\mu\text{m}^2$ is the effective mode area, $\gamma=\omega_0 n_2/(cA_{\text{eff}})$, $\tau=12$ ns is the phonon lifetime, and $P_T(z,t)=\sum_{j=0}^{2}\left[P_j^+(z,t)+P_j^-(z,t)\right]$. We assume that $E_j^{\pm}(z,t)$ and $M_j^{\pm}(z,t)$ are normalized such that $P_j^{\pm}(z,t)=|\tilde{A}_j^{\pm}(z,t)|^2$ and $\tilde{Q}_j^{\pm}(z,t)$ are measured in dimensions of Watts and that $P_j^{\pm}(z)$ represents the optical power[5]. $\tilde{Q}_j^{\pm}(z,t)$ is related to the slowly varying complex envelope of the density variation[6] $\rho_j^{\pm}[\text{kg/m}^3]$ with $\tilde{Q}_j^{\pm}=\rho_j^{\pm}\times(v_a^2 A_{\text{eff}}c/ip_{12}\Omega_B\tau n^3)$, where $v_a=\Omega_B\lambda/(4\pi n)$ is the velocity of the acoustic wave and $p_{12}=0.266$ is the elasto-optic coefficient of $As_2Se_3$[6]. The dispersion of the effective index[7] $n$ and the Brillouin frequency shift $\Omega_B$ were assumed to be negligible. In equation (S6) and (S7) the spatial derivatives of $\tilde{Q}_1^{\pm}(z,t)$ and $\tilde{Q}_2^{\pm}(z,t)$ describing acoustic-wave propagation are neglected because the acoustic waves are strongly damped[5]. The Langevin noise terms[8] $f_j^{\pm}(z,t)$ are Gaussian random variables with zero mean $\langle f_j^{\pm}(z,t)\rangle=0$, which are $\delta$-correlated in the sense $\langle f_k^{\pm}(z_1,t_1)f_l^{\pm*}(z_2,t_2)\rangle=\psi\delta(z_1-z_2)\delta(t_1-t_2)\delta_{kl}$, where $f_k^+(z,t)$ and $f_k^-(z,t)$ are also uncorrelated. In the normalization used here, the strength parameter $\psi \approx 2k_B T \rho_0 v_a^2 A_{eff} c^2 / (\tau_B^3 p_{12}^2 n^6 \Omega_B^2)$, where $\rho_0 = 4640 \text{ kg/m}^3$ is the density of As$_2$Se$_3$[7], $k_B$ is the Bolzmann-constant and $T = 293 \text{ K}$ is the temperature of the resonator.

All fields set to zero for t<0 and the SBS process is initiated by an input pump field $E_{in}(t) = \sqrt{P_{in}(t)} \exp(-i\omega_0 t)$, which is coupled into the fibre at $z = 0$ and $t \geq 0$. The boundary conditions of the system are given by

$$\tilde{A}_0^+(0,t) = \sqrt{R}\tilde{A}_0^-(0,t) + \sqrt{P_{in}(t)}$$
$$\tilde{A}_j^+(0,t) = \sqrt{R}\tilde{A}_j^-(0,t); \text{ for } j = 1, 2. \quad \text{(S8)}$$
$$\tilde{A}_j^-(L,t) = \sqrt{R}\tilde{A}_j^+(L,t)e^{i\Delta\varphi_j}; \text{ for } j = 0, 1, 2.$$

$\Delta\varphi_j = 2\tilde{\omega}_j nL/c$ are linear phase shifts of pump and Stokes waves due to transit through the cavity. It is convenient to describe these linear phase shifts in the form $\Delta\varphi_j = \Delta\varphi_0 - j\beta$, (j=0,1,2), where $\Delta\varphi_0 = 2\tilde{\omega}_0 nL/c$ is the linear phase-shift of the pump per roundtrip through the cavity and $\beta = 2\Omega_B nL/c$ is the difference of the linear phase shifts of the two adjacent comb components (without Stokes detuning) per roundtrip through the cavity[4].

**Steady state equations**

Here we derive the steady state equations for the Powers $P_j^\pm(z)$ and phases $\varphi_j^\pm(z)$ of the optical waves allowing stable Stokes detuning of first and second-order Stokes waves. We start by inserting

$$\tilde{A}_0^\pm(z,t) = A_0^\pm(z),$$
$$\tilde{A}_j^\pm(z,t) = A_j^\pm(z)e^{-i\Delta\omega_j(t \mp nz/c)}, j = 1, 2, \quad \text{(S9)}$$
$$\tilde{Q}_j^\pm(z,t) = Q_j^\pm(z)e^{i\Delta\omega_j(t \pm nz/c)}, j = 1, 2,$$

into equations (S3)-(S7) and setting $\partial A_j^\pm(z)/\partial t = \partial Q_j^\pm(z)/\partial t = f_j^\pm = 0$. Equations (S6) and (S7) can then be solved analytically and inserted in the equations (S3) - (S5). Inserting the substitution $A_j^\pm(z) = \sqrt{P_j^\pm(z)} \exp(i\varphi_j^\pm(z))$ in the resulting equations and separating real and imaginary parts yields

$$\pm \frac{\partial P_0^\pm}{\partial z} = -\alpha P_0^\pm - \frac{g_B}{A_{eff}(1+\xi_1^2)} P_1^\mp P_0^\pm + \left(2\gamma P_1^\pm \sqrt{P_2^\pm P_0^\pm}\right)\sin\vartheta^\pm \quad (S10)$$

$$\pm \frac{\partial P_1^\pm}{\partial z} = -\alpha P_1^\pm + \frac{g_B}{A_{eff}(1+\xi_1^2)} P_0^\mp P_1^\pm - \frac{g_B}{A_{eff}(1+\xi_2^2)} P_2^\mp P_1^\pm - \left(4\gamma P_1^\pm \sqrt{P_0^\pm P_2^\pm}\right)\sin\vartheta^\pm \quad (S11)$$

$$\pm \frac{\partial P_2^\pm}{\partial z} = -\alpha P_2^\pm + \frac{g_B}{A_{eff}(1+\xi_2^2)} P_1^\mp P_2^\pm + \left(2\gamma P_1^\pm \sqrt{P_0^\pm P_2^\pm}\right)\sin\vartheta^\pm \quad (S12)$$

$$\pm \frac{\partial \varphi_0^\pm}{\partial z} = \frac{g_B \xi_1}{2(1+\xi_1^2)A_{eff}} P_1^\mp + \gamma\left(2P_T - P_0^\pm\right) + \gamma P_1^\pm \sqrt{P_2^\pm / P_0^\pm} \cos\vartheta^\pm \quad (S13)$$

$$\pm \frac{\partial \varphi_1^\pm}{\partial z} = \frac{g_B \xi_1 P_0^\mp}{2(1+\xi_1^2)A_{eff}} + \frac{g_B \xi_2 P_2^\mp}{2(1+\xi_2^2)A_{eff}} + \gamma\left(2P_T - P_1^\pm\right) + 2\gamma \sqrt{P_0^\pm P_2^\pm} \cos\vartheta^\pm, \quad (S14)$$

$$\pm \frac{\partial \varphi_2^\pm}{\partial z} = \frac{g_B \xi_2 P_1^\mp}{2(1+\xi_2^2)A_{eff}} + \gamma\left(2P_T - P_2^\pm\right) + \gamma P_1^\pm \sqrt{P_0^\pm / P_2^\pm} \cos\vartheta^\pm, \quad (S15)$$

where $\xi_1 = -2\tau\Delta\omega_1$, $\xi_2 = -2\tau\Delta\omega_2$, and the phase parameter $\vartheta^\pm(z,t)$ is defined in the same way as in the main document in equation (5).